\documentclass[]{IEEEapm}

\jmonth{}
\pubyear{2022}

\usepackage{color}
\usepackage{pdfpages}

\begin{document}

\title{Using Time-Varying Systems to Challenge Fundamental Limitations in Electromagnetics}

\author{Zeki~Hayran, Francesco~Monticone}

\affil{School of Electrical and Computer Engineering, Cornell University, Ithaca, New York 14853, USA}  

\maketitle

\markboth{ }{ }

\begin{receivedinfo}
 
\end{receivedinfo}

\begin{abstract}
Time-varying systems open intriguing opportunities to explore novel approaches in the design of efficient electromagnetic devices. While such explorations date back to more than half a century ago, recent years have experienced a renewed and increased interest into the design of dynamic electromagnetic systems. This resurgence has been partly fueled by the desire to surpass the performance of conventional devices, and to enable systems that can challenge various well-established performance bounds, such as the Bode-Fano limit, the Chu limit, and others. Here, we overview this emerging research area and provide a concise and systematic summary of the most relevant applications for which the relevant performance bounds can be overcome through time-varying elements. In addition to enabling devices with superior performance metrics, such research endeavors may open entirely new opportunities and offer insight towards future electromagnetic and optical technologies.
\end{abstract}


\section{Introduction}

The study of electromagnetic wave interactions with time-varying media dates back to more than half a century ago, when several pioneering works \cite{morgenthaler1958velocity, simon1960action, cassedy1963dispersion, cassedy1965waves, holberg1966parametric, cassedy1967dispersion, auld1968signal, felsen1970wave, fante1971transmission, chu1972wave} revealed the rich physics arising due to such interactions. Although in the subsequent decades several other works have further explored the implications of wave propagation in dynamic media \cite{yablonovitch1973spectral, stepanov1976dielectric, kalluri1992frequency, stepanov1993waves}, it was not until recently when the field gained considerable momentum \cite{caloz2019spacetime} partly owing to the improved fabrication and characterization techniques in various frequency regimes \cite{shaltout2019spatiotemporal, taravati2020space, reshef2019nonlinear}. Consequently, it was soon discovered that time-dependent systems can enable many interesting and novel phenomena, such as interband photonic transitions \cite{winn1999interband,yu2009complete}, nonreciprocal systems \cite{sounas2017non, williamson2020integrated}, efficient frequency conversion \cite{notomi2006wavelength, preble2007changing, zhou2020broadband, khurgin2020adiabatic}, temporal aiming \cite{pacheco2020temporal}, Fresnel drag \cite{huidobro2019fresnel}, negative extinction \cite{shcherbakov2019time}, entangled photon generation \cite{cirone1997photon, sloan2021casimir}, parametric oscillations \cite{lee2021parametric}, synthetic dimensions \cite{yuan2021synthetic}, topological phase transitions \cite{ozawa2019topological}, nonreciprocal gain \cite{koutserimpas2018nonreciprocal}, among others \cite{caloz2019spacetime2}. Several recent publications also further investigated the fundamental aspects of wave propagation in time-varying media, such as Refs. \cite{shvartsburg2005optics, biancalana2007dynamics, xiao2011spectral, bakunov2014reflection, plansinis2015temporal, plansinis2016spectral}, thus providing a better understanding of the physical implications of these systems.

Apart from enabling novel effects, time-varying materials and components may also be used to improve a given device to achieve better performance in terms of operational bandwidth, size, efficiency of the relevant process, and other metrics \cite{caloz2019spacetime2}. In some cases, such improvements may even exceed what is normally allowed by well-established performance bounds, as conventional physical bounds and fundamental limits in electromagnetics are typically derived under the assumption of linearity and time-invariance (LTI systems). Indeed, as mentioned above, time-variance in the system properties may enable various functionalities not present in a conventional setting, such as parametric gain \cite{koutserimpas2018nonreciprocal} or spectral changes \cite{winn1999interband, notomi2006wavelength, preble2007changing, zhou2020broadband, khurgin2020adiabatic,hayran2021spectral}, which are not accounted for when deriving bounds for LTI systems, and can therefore be exploited to bypass such physical limits. Consequently, time-varying materials and components provide exceptional opportunities to realize superior devices not restricted by various conventional limits.

\section{Applications}

In this section we provide a concise overview of the most relevant applications of time-varying electromagnetic systems to overcome various physical bounds of conventional LTI systems. In this regard, this section is divided into various subsections, each of which focuses on a specific application or research topic, with the goal to systematically guide the reader through this emerging research area. We will not discuss some major research directions that also seek to overcome various constraints of conventional systems and broaden their functionality, such as the possibility to break reciprocity \cite{sounas2017non, williamson2020integrated} or create synthetic dimensions \cite{yuan2021synthetic} based on time modulation, since these topics are already covered in detail in the literature. Instead, we will focus on certain topics of particular interest to the IEEE Antennas and Propagation community, including how time-varying systems impact the fundamental limitations of antennas \cite{hansen1981fundamental}, broadband impedance-matching networks \cite{fano1950theoretical}, and radar absorbers \cite{rozanov2000ultimate}.

\subsection{Impedance Matching}
\label{section_impedance}

Efficient power transfer from a source to a load with minimal reflections is arguably one of the most important tasks in electromagnetic engineering. Thus, when designing impedance-matching networks to minimize reflections, it becomes essential to identify the limitations and the optimal configurations of the system. In this regard, a matching-bandwidth limit was introduced for linear, time-invariant, passive systems by Bode in 1945 \cite{bode1945network}, which was later generalized by Fano for arbitrary load impedances \cite{fano1950theoretical}. The resulting Bode-Fano limit provides a theoretical upper bound on the bandwidth over which a certain reflection reduction can be attained, independent of the complexity of the employed matching network (only assumed to be reactive), and it is therefore a very powerful tool to assess the maximum possible bandwidth of various electromagnetic systems, such as antennas \cite{gustafsson2006bandwidth} or optical invisibility cloaks \cite{monticone2016invisibility}. To overcome this limitation, one strategy that has been explored in the literature is to break the underlying assumption of passivity. Indeed, several proposals utilizing active ``non-Foster'' elements have shown that the bandwidth of small antennas \cite{sussman2009non} and cloaking devices \cite{hrabar2010towards, chen2013broadening, soric2014wideband, chen2019active} can be extended beyond what would be possible with a passive system. However, active elements typically increase the noise level of the system \cite{jacob2016gain} and can lead to instabilities (unbounded oscillations in time) \cite{chen2019active, ugarte2012stability, abdelrahman2021physical}, thereby disrupting the desired functionality. 

\begin{figure}[t]
\centering
\includegraphics[width=1.0\linewidth]{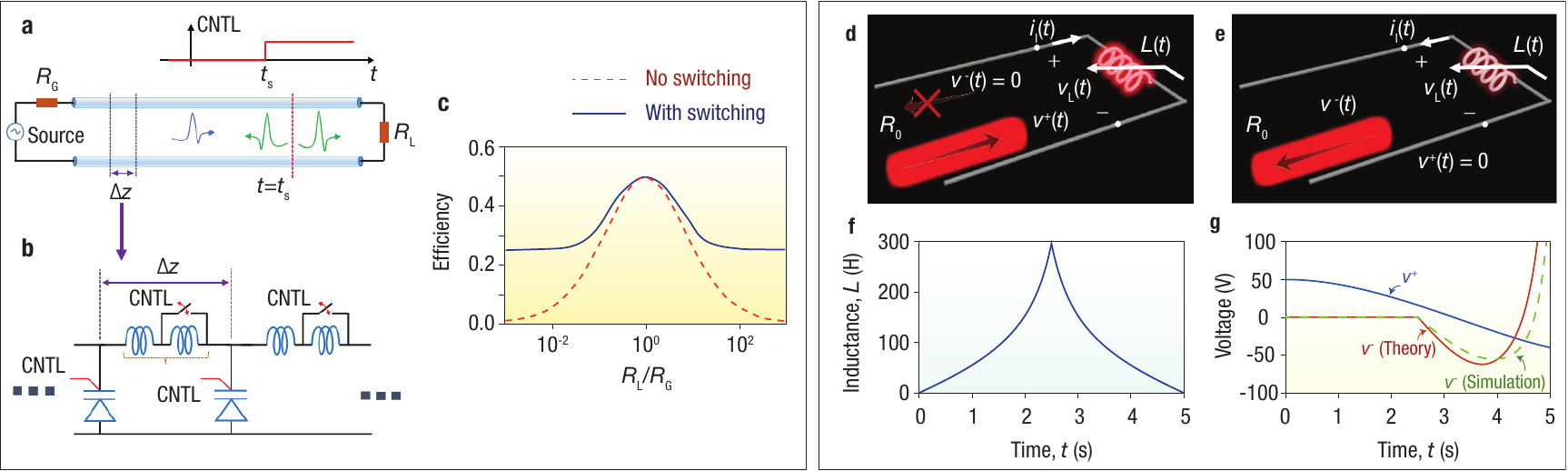}
\caption{(a) A temporally switched transmission line can be employed to match a source impedance $R\textsubscript{G}$ to a load impedance $R\textsubscript{L}$ surpassing the Bode-Fano limit. (b) Such a switched matching line can be realized with parallel varactor diodes and banks of series inductors, whose properties are changed abruptly in time at $t = t\textsubscript{s}$. (c) The maximum transmission efficiency for various $R\textsubscript{L}/R\textsubscript{G}$ values shows that the temporally switched case (solid blue) provides better performance compared to the non-switched time-invariant case (dashed red line) especially for large source-load impedance contrast values. (d),(e) A transmission line terminated with a time-varying inductance $L(t)$. (d) Using a suitable time modulation, the inductor can accumulate energy without any reflections. (e) By time-reversing the temporal modulation in (d) the energy trapped by the inductor can be released. (f) Time modulation profile of the inductance, comprising both the trapping (before $t=2.5$ s) and releasing regimes (after $t=2.5$ s). (g) Temporal variation of the incident signal $V^+$ and the reflected signal $V^-$ at the load position (obtained theoretically and through simulations) for a load inductance temporally modulated as in panel (f). Panels (a-c) are reproduced with permission from Ref. \cite{shlivinski2018beyond}, American Physical Society. Panels (d-g) are reproduced with permission from Ref. \cite{mirmoosa2019time}, American Physical Society.}
\label{impedance}
\end{figure}

An alternative option to go beyond the Bode-Fano limit is to break the assumption of time-invariance, i.e., allow temporal variation in the system properties. In this context, Ref. \cite{wang2007time2} employed time-varying ``switched'' matching networks to achieve a better bandwidth performance compared to their time-invariant counterparts. More recently, Ref. \cite{shlivinski2018beyond} provided another interesting method, which involves switching the system parameters while a broadband signal (a pulse) is within the matching network and before it reaches the load, as shown in Figs. \ref{impedance}(a) and \ref{impedance}(b). The temporal switching would generally lead to reflected waves due to the temporal discontinuity (temporal interface) \cite{mendoncca2002time}; however, it was shown that the matching efficiency (defined as the ratio between the power dissipated on the load, and the maximal available power from the source plus the total power provided by the modulation) can still be maximized over a broad bandwidth by optimizing several parameters, including the characteristic impedance of the matching transmission line and the phase velocity of the pulse within the matching line before and after the switching event \cite{shlivinski2018beyond}. Since temporal switching increases the number of free parameters that can be tuned compared to the time-invariant case, one may achieve performance metrics that are otherwise out of reach in a conventional time-invariant setting, especially for large load-source impedance contrast values (see Fig. \ref{impedance}(c)) \cite{shlivinski2018beyond}. This proposal was further extended from ``hard'' (abrupt) to ``soft'' temporal switching in Ref. \cite{hadad2020soft}, which could be more feasible to implement in practice. An adverse byproduct of the temporal switching operation is the spectral compression or broadening of the pulse, which necessitates the use of additional analog or digital devices to restore the original pulse spectrum and, thus, may complicate the overall matching system. Moreover, this method is feasible only for short-time signals (as quasi-monochromatic signals would require impractically long matching lines), and requires knowledge that the pulse is present within the time-switched matching line (which implies the presence of some fast detection mechanism able to trigger the switching event). Even with these limitations, this approach shows a realistic path towards broadband matching networks beyond the Bode-Fano constraints.

For the quasi-monochromatic regime, on the other hand, a transmission line system where the load impedance itself is varied in time has been used to impedance match a time-harmonic source to a purely reactive load and to accumulate and release energy on demand (see Fig. \ref{impedance}(d-g)) \cite{mirmoosa2019time}. Specifically, it was shown that, while a time-invariant reactive load would fully reflect an incident wave, a time-modulated reactance can emulate a resistance and, can, therefore act as a ``virtual'' absorber \cite{mirmoosa2019time}. An intuitive explanation of this effect is that temporally modulating the reactance with a suitable tangent function is equivalent to a short-circuited transmission line with the shorted end moving away from the source with the same velocity as the phase velocity of the incident signal \cite{mirmoosa2019time}. Therefore, the incident wave does not reflect back since it never appears to ``reach'' the reflecting termination. Consequently, energy from a time-harmonic source can be trapped with no reflections and released on demand (see Figs. \ref{impedance}(f) and \ref{impedance}(g)). 

Lastly, another interesting application of temporal modulation in the context of impedance matching is related to the problem of capturing and trapping an incident electromagnetic wave into a \emph{lossless} cavity, with no reflections. This can be achieved by temporally modulating the radiation leakage from the cavity, such that it perfectly cancels the direct reflection from the cavity at all time instants \cite{sounas2020virtual}. In this regard, we have also shown that by temporally modulating the shell of a compact core-shell scatterer (acting as an open cavity supporting a nonradiating mode), an incident broadband pulse can be captured within the cavity with no reflection, effectively forcing a broadband signal into a narrowband resonator with ideal efficiency (this may also be interpreted as spectrum compression through temporal modulation with the additional constraint of zero reflections) \cite{hayran2021capturing}. Further details are provided in the Supplementary Material.

\subsection{Antenna Performance Enhancement}

Antenna miniaturization is another major research area within the applied electromagnetics community with many potential benefits for various applications, such as communication systems, sensor networks, implanted medical devices, and navigation systems. While it is generally possible to reduce the size of an antenna down to subwavelength dimensions for a given design frequency, this typically comes at the price of reduced bandwidth and radiation efficiency. In this regard, the fundamental trade-offs between size and bandwidth for electrically small antennas were investigated more than half a century ago in several seminal papers by Wheeler \cite{wheeler1947fundamental}, Chu \cite{chu1948physical}, and later by Harrington \cite{harrington1960effect}, resulting in, as it is known today, the Chu-Harrington limit \cite{yarman2008design}. The limit as derived by Chu is based on an equivalent circuit model for the radial wave impedance of the lowest-order spherical wave radiated by an antenna enclosed within a spherical volume. Regardless of the antenna structure and LTI material, this equivalent circuit for spherical waves becomes more reactive as the size of the spherical volume is reduced, leading to a higher Q factor (consistent with the fact that, for example, a short wire dipole has a strongly reactive input impedance). Then, as recognized by Chu, while the inverse of the Q factor may be interpreted as the fractional bandwidth of the antenna, a more accurate definition is in terms of the tolerable reflection coefficient over the operating band, and therefore the antenna bandwidth is limited by the Bode-Fano limit (see Section \ref{section_impedance}) \cite{lopez2013historical} applied to Chu's equivalent circuit. In other words, these results define a tradeoff between antenna bandwidth and realized gain (gain taking into account the efficiency associated with reflection losses), and quantify how the product between bandwidth and realized gain is decreased by miniaturization. These analyses were later refined by other researchers (e.g. \cite{hansen1981fundamental, mclean1996re, gustafsson2007physical,yaghjian2018overcoming}), further underlying the fact that for a linear time-invariant passive antenna one cannot simultaneously minimize size and maximize bandwidth and efficiency.

Following Chu's work, antenna engineers realized early the need for active (i.e., non-passive) designs, nonlinear materials, and time-varying components to increase the bandwidth of wireless systems based on electrically small antennas. In this context, for example, nonlinear ferrite materials were exploited to realize time-varying inductors, which were used to overcome the limitations of small antennas in VLF transmitters by shifting the carrier frequency \cite{jacob1954keying}. Subsequent works (see, for instance, Refs. \cite{wolff1957high, galejs1963switching, yao2004radiating, wang2007time, wang2010analysis}) further demonstrated the unique opportunities enabled by time-varying ``switched'' elements to design antennas with improved characteristics.

\begin{figure}[t]
\centering
\includegraphics[width=0.9\linewidth]{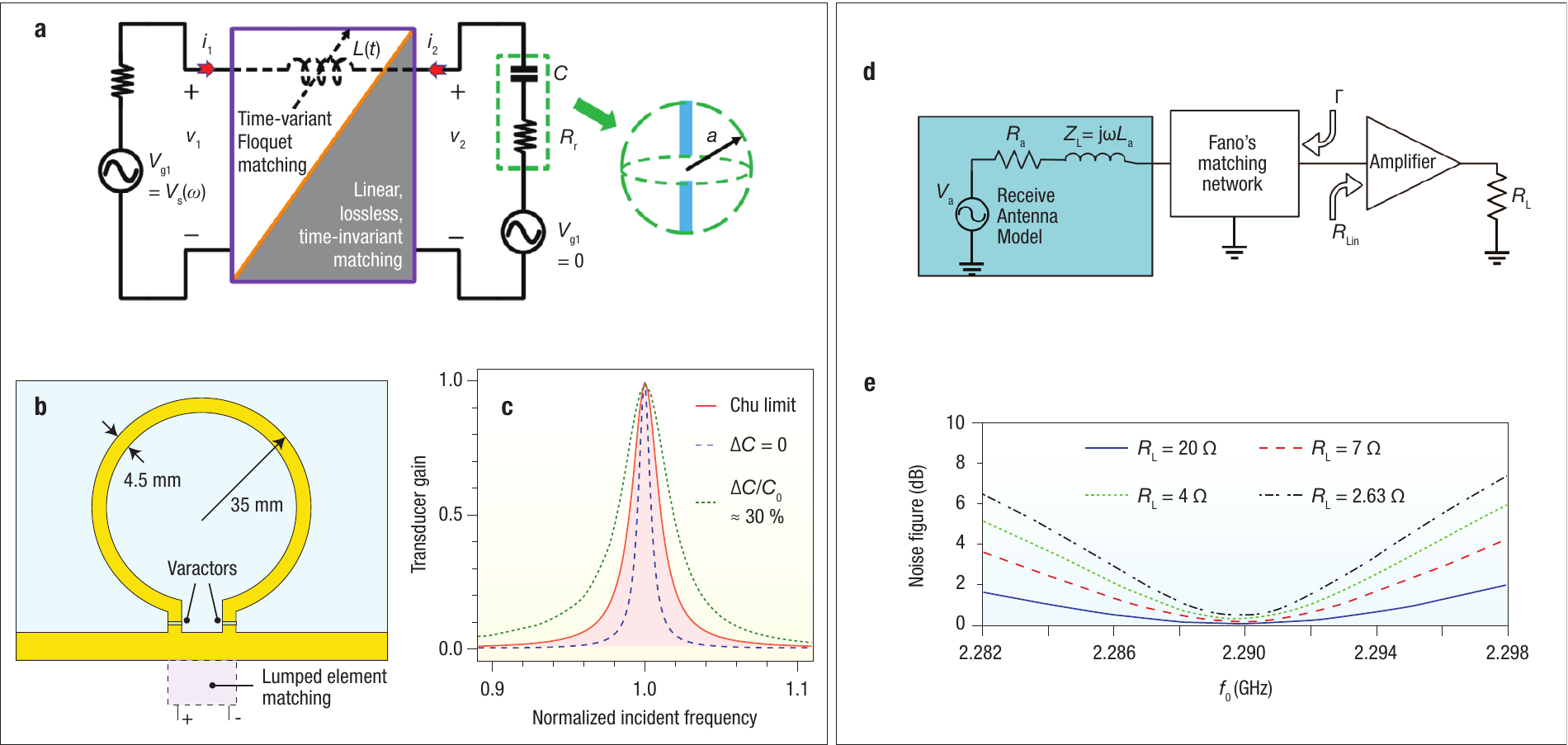}
\caption{(a) A ``Floquet'' matching network with periodically time-varying elements can be employed to increase the efficiency-bandwidth product of an electrically small antenna by imparting parametric gain to the system. (b) An electrically small loop antenna (at the design frequency of 300 MHz) matched with periodically-driven variable capacitors (varactors). (c) Transducer gain profiles (efficiency taking into account mismatch losses) obtained through numerical simulations show that the bandwidth of the time-modulated antenna (green dotted line) can exceed the maximum bandwidth predicted by the Bode-Fano limit applied to the Chu-Harrington equivalent circuit (red solid line). The time-invariant antenna (blue dashed line), on the other hand, is bounded by this limit. The incident frequency is normalized to the antenna design frequency. $\Delta C$ denotes the capacitance modulation amplitude, while $C_0 $ ($\approx$3 pF) is the static capacitance. (d) Schematics for a different example of (receiving) antenna with enhanced efficiency-bandwidth product. By deliberately inducing a mismatch between the antenna and the load, hence increasing reflection losses, the antenna bandwidth can be widened. The decrease in efficiency associated with mismatch losses can then be compensated through a parametric amplifier. (e) The inherently low-loss nature of parametric amplification can help bring the noise figure down while still maintaining a high bandwidth enhancement. Panels (a-c) are reproduced with permission from Ref. \cite{li2019beyond}, American Physical Society. Panels (d,e) are reproduced with permission from Ref. \cite{loghmannia2021broadband}, IEEE.}
\label{chu}
\end{figure}

More recently, with the renewed surge of interest in temporal modulation in electromagnetic systems, research into electrically small time-varying antennas has been reinvigorated. For example, Ref. \cite{manteghi2009antenna} confirmed that the antenna bandwidth can be increased significantly through impedance modulation. Another more recent work employed ``Floquet'' impedance matching \cite{li2019beyond} (see Fig. \ref{chu}) to overcome the Chu-Harrington limit by imparting parametric gain into the system \cite{lee2021parametric} through a periodic modulation of the matching network components at twice the frequency of the antenna resonant frequency. A realistic simulation of the proposed design was also demonstrated based on a small loop antenna (see Figs. \ref{chu}(b)) that includes periodically modulated variable capacitances (varactors). Through numerical calculations (see Fig. \ref{chu}(c)), it was shown that the bandwidth-efficiency product can be significantly enhanced owing to the temporal modulation compared to the time-invariant case, and can even surpass the predictions of the Bode-Fano limit applied to the Chu-Harrington equivalent circuit, which is a rather remarkable result given the fact that parametric phenomena are typically narrowband. A possible drawback of schemes of this type is that the parametric gain may potentially lead to instabilities. However, unlike antenna systems based on active non-Foster elements \cite{zhu2012broad,church2014uhf,shi2019improved}, the dispersion of the parametric gain can be more easily controlled through the modulation properties to bring the system into the stable regime. Moreover, it was also pointed out that deliberately inducing an impedance mismatch by modifying the source impedance can act as a negative feedback to help ensure stability and increase the operational bandwidth at the same time \cite{li2019beyond}. Indeed, a stable parametric enhancement effect was later experimentally demonstrated in Ref. \cite{mekawy2021parametric} in a fabricated time-varying loop antenna operating at RF frequencies. 
In this context, other recent studies also investigated the use of time-varying parametric amplifiers with the aim of broadening the bandwidth of electrically small antennas \cite{loghmannia2020low, loghmannia2021broadband} (see Figs. \ref{chu}(d,e)). Similar to Refs. \cite{li2019beyond, mekawy2021parametric}, this was achieved by first trading realized gain with bandwidth, in line with the Bode-Fano limit, by detuning the system from the ideal impedance-matched condition, hence increasing reflections, and subsequently using parametric up-converters based on time-varying reactive elements to compensate for these mismatch losses. Different from Refs. \cite{li2019beyond, mekawy2021parametric}, however, in Refs. \cite{loghmannia2020low, loghmannia2021broadband} the authors placed a particular focus on assessing the noise properties of the system and studied a specific circuit configuration in which a receiving antenna is connected to an amplifier (Fig. \ref{chu}(d)) with real input impedance several times greater than the real part of the input impedance of the antenna, as this allows achieving an optimal trade-off between increased bandwidth and added noise (Fig. \ref{chu}(e)).

\subsection{Electromagnetic Absorption}

\begin{figure}[t]
\centering
\includegraphics[width=1.0\linewidth]{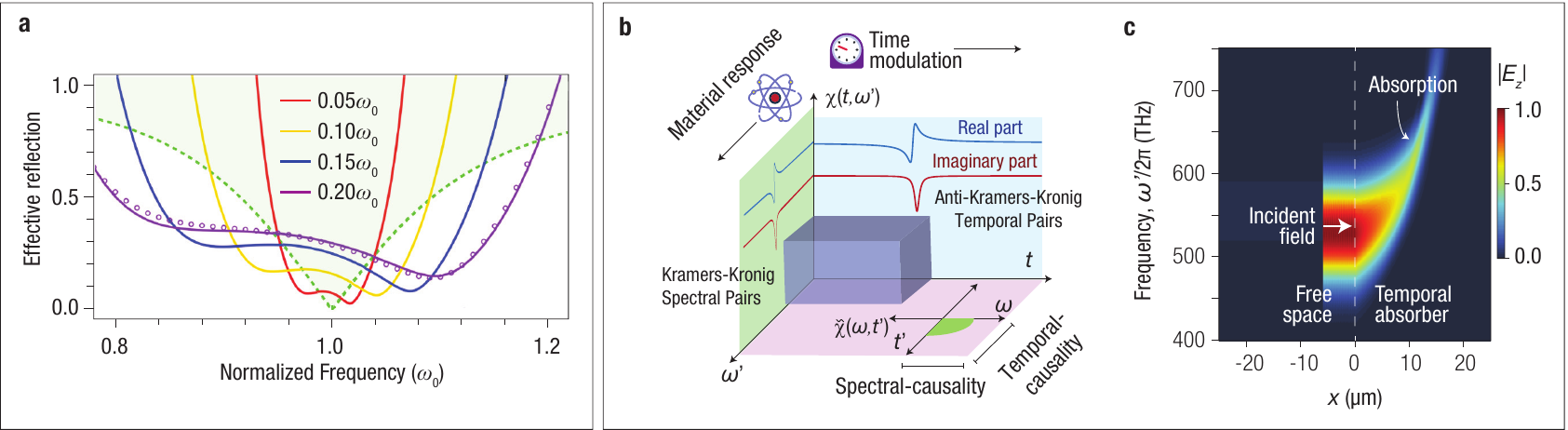}
\caption{(a) An electromagnetic absorber with a time-switched permittivity can be employed to overcome the Rozanov limit on broadband electromagnetic absorption (shown by the green dashed line). Different reflection spectra are shown, corresponding to incident Gaussian pulses with different widths. (b) Alternatively, one may temporally modulate the complex permittivity (both its real and imaginary parts simultaneously) such that the modulation spectrum becomes one-sided (or ``spectrally casual''), which implies that frequency down-conversion and back-reflections due to the temporal modulation are prohibited over a broad bandwidth. (c) Such a complex modulation may potentially increase the bandwidth of electromagnetic absorbers, as in the time-switched case, but without inducing any reflection within or outside the operational bandwidth (as seen in the figure, the field is fully transmitted, frequency up-shifted, and absorbed, with no reflections, at any frequency). Panel (a) is reproduced with permission from Ref. \cite{li2021temporal}, Optical Society of America. Panels (b) and (c) are reproduced with permission from Ref. \cite{hayran2021spectral}, Optical Society of America.}
\label{absorption}
\end{figure}

Electromagnetic wave absorption plays a crucial role in various electromagnetic and optical systems, ranging from solar energy harvesting to scattering minimization for radar applications or anechoic chamber measurements.  
A key limitation of conventional electromagnetic absorbers is the typical trade-off between the absorption bandwidth, i.e., the bandwidth over which a certain level of absorption can be maintained, and the thickness of the absorbing material or structure. Within this context, using the analytical properties of the reflection coefficient, Rozanov \cite{rozanov2000ultimate} showed that this trade-off is in fact fundamental, and there exists a ultimate thickness to bandwidth ratio $\Delta\lambda/d$ for passive time-invariant absorbers backed by a perfect conductor. Specifically, for the case of a nonmagnetic absorber, the upper limit for this ratio is $\Delta\lambda/d<16/|\ln\rho_0|$, where $\rho_0$ is the maximum tolerable reflection within the operating bandwidth. Clearly, the absorption bandwidth can therefore be widened by increasing the thickness of the absorber or the tolerable reflection. Prominent examples of broadband absorbers include cascaded \cite{he2015ultra, yang2021visible} or adiabatically tapered \cite{cui2012ultrabroadband, liang2013metamaterial, hayran2017stopped} lossy structures, which typically have thicknesses on the order of several wavelengths. However, for many applications, such as radar absorption or solar energy harvesting, creating the thinnest possible absorber with the widest possible bandwidth is highly desirable, which motivates the significant interest in overcoming the ``Rozanov bound''. In this direction, it was shown in Ref. \cite{li2021temporal} that temporally switching the relative permittivity of the absorbing material can be used to extend the bandwidth of thin absorbers beyond what is possible with time-invariant structures (see Fig. \ref{absorption}(a)), similar to the use of temporally switched transmission lines for broadband impedance matching as discussed in Section \ref{section_impedance}. An important drawback of these approaches is that, while reflections are reduced within the bandwidth of the incident wave, significant reflections may be induced outside this frequency range as the temporal switching can lead to strong frequency conversion. While this may not pose a problem for certain narrowband systems, it would become an issue in the context of, for example, broadband radar and stealth technology, as the target may become even more easily detectable due to the increased reflections outside the original signal bandwidth, which could then be picked up by sufficiently broadband or multi-band detectors. To overcome this problem, a possible strategy is to modulate both the real and imaginary part of the permittivity in such a way that the modulation spectrum becomes ``unilateral'' (or ``spectrally causal''), as we theoretically demonstrated in a recent publication \cite{hayran2021spectral}. Such an approach ensures that back-reflections due to the temporal modulations are prohibited over a very wide frequency range and, therefore, the incident wave is perfectly absorbed with no reflections over wide bandwidths (see Fig. \ref{absorption}(b),(c)). Along similar lines, a recent work theoretically demonstrated, through a Green's function analysis and numerical optimizations, that a suitable switching of both the conductivity and permittivity leads to a reduction of reflection over the whole frequency spectrum \cite{firestein2022absorption}. As an alternative strategy, a recent experimental work \cite{yang2022broadband} also demonstrated broadband absorption by creating an energy-trap through switched electronic components triggered by the pulse entering the absorbing region.
While the strategies described above work for short broadband pulses and are based on non-periodic temporal modulations (e.g., ``switching''), perfect absorption for quasi-monochromatic signals has also been realized by employing periodic temporal modulations \cite{suwunnarat2019dynamically}. Specifically, it was shown that perfect absorption can be achieved even with infinitesimal losses owing to coherent multichannel illumination with Floquet time-periodic driving schemes. A single-channel coherent perfect absorber based on a time-Floquet engineered reflector without any physical mirror (ground plane) was also demonstrated, which might be useful in certain application scenarios where metallic or Bragg mirrors are not available.

\subsection{Other Applications}

Our list of applications that can benefit from time-varying systems is certainly not exhaustive. In the Supplementary Material, we provide several other application examples that might be of interest to readers also from different backgrounds, such as optics and photonics.

\section{Outlook}
Although the study of time-varying electromagnetic systems dates back more than half a century, recent years have seen a surge of renewed interest in this topic, largely motivated by the need to realize novel electromagnetic systems that are superior to conventional LTI ones for various applications, from high-speed full-duplex communication systems, to broadband stealth technology, to high-efficiency energy harvesting and energy transfer. However, while recent progress in this field has been promising, we believe several challenges should be taken into consideration and carefully addressed before such systems can be adopted into practical use.

Since a time-varying system requires external power to operate, such active systems might become susceptible to instabilities (unbounded oscillations) that can disrupt the desired functionality. Thus, especially for systems involving parametric amplification, a careful analysis is essential to assess and ensure stability \cite{li2019beyond}. Moreover, the additional energy pumped into the system \cite{secondo2020absorptive} should be carefully taken into account when determining the figure of merits (such as power efficiency) of the design \cite{jayathurathnage2021time} and also to assess whether the external power required by the temporal modulation is beyond practical reach and/or may damage the system \cite{hayran2022omega}.

Frequency dispersion is another fundamental property of physical materials, ultimately rooted in the implications of the causality principle. Hence, any realistic system involves, to some degree, frequency-variations in the material constitutive parameters. However, especially for numerical studies involving time-varying structures, such dispersion effects are often ignored and the time-variation is assumed to be independent of frequency \cite{yu2009complete}. As a result, any realistic implementation of such proposals will necessarily experience performance deterioration due to frequency dispersion of the material properties. We note that several recent works \cite{hayran2022omega, mirmoosa2022dipole, solis2021functional, bakunov2021constitutive} have now started studying material dispersion effects more systematically in the context of time-varying systems. However, we believe further investigations in this area are important to enhance our understanding of time-varying materials and may lead to improved designs and more accurate performance predictions.

For schemes requiring synchronization between the incident wave and the temporal variation, another significant challenge is that fast detectors and switching mechanisms are essential to trigger the modulation in the presence of the incident wave. Such additional components might complicate the design and reduce its practical applicability. Hence, more research is needed to integrate such components into time-varying systems in a practically useful manner.

Dissipation, unless desired to realize absorbers, is another challenge that needs to be addressed for various practical applications that involve time-varying systems. For instance, epsilon-near-zero materials have recently become the subject of increasing interest to realize deep and fast modulations at optical frequencies, as they can be efficiently controlled through external optical pumps \cite{reshef2019nonlinear}. However, such materials still suffer from high losses around their epsilon-near-zero wavelengths. Hence, the search for better epsilon-near-zero materials with lower losses is an important area of research (note that, theoretically, nothing prevents the existence of an epsilon-near-zero material that is dissipationless, albeit frequency dispersive, within a certain frequency window). Alternatively, one could exploit (rather than eliminate or compensate) the losses in such materials to realize time-dependent loss profiles that accompany the time-varying refractive-index profiles. In this context, we recently provided a general strategy to exploit the synergy between these two time-modulation profiles for various applications \cite{hayran2021spectral}. Investigations along these directions may also enhance our understanding of wave dynamics in complex (``non-Hermitian'') time-varying media.

Despite the challenges outlined here, we believe that the use of time-varying materials and components to overcome the limitations of conventional LTI electromagnetic/photonic systems is an exciting and promising research area that might lead to novel and superior devices and, as more research is conducted, a deeper understanding of the actual fundamental limits in electromagnetics. Spatially-varying engineered structures such as frequency selective surfaces, diffraction gratings, photonic crystals, and metamaterials have revolutionized the field of electromagnetics in recent decades. Along the same vein, it would be safe to say that the new insight and the new tools provided by the field of time-varying electromagnetic systems will enable further advances and possible revolutions in the next decades.

\section*{Acknowledgements}
The authors acknowledge support from the Air Force Office of Scientific Research with Grant No. FA9550-19-1-0043 through Dr. Arje Nachman.

\bibliographystyle{IEEEtran}
\bibliography{Limits_Time_Varying.bbl}

\setboolean{@twoside}{false}


\section*{Supplementary material}

# Supplementary Material: Using Time-Varying Systems to Challenge Fundamental Limitations in Electromagnetics

**Zeki Hayran, Francesco Monticone**

*School of Electrical and Computer Engineering, Cornell University, Ithaca, New York 14853, USA*

## 1. Other Applications

Given their design flexibility, time-varying electromagnetic systems have provided a promising way to overcome conventional performance limitations in many other application areas not discussed in the main article. In this Supplementary Material, we will mention several other examples to illustrate how far reaching the field of time-varying electromagnetics can be.

### *1.1. Storing and Delaying Electromagnetic Pulses*

The ability to dramatically slow down an electromagnetic pulse through delay lines or buffers has broad implications for many communication systems. In this regard, a delay line that reduces the group velocity of the incoming pulse over a broad bandwidth $\Delta\omega$ is typically desired to yield a large storage or buffering capacity. However, a large bandwidth often comes at the price of a reduced delay time $\Delta T$ [1]. This trade-off between bandwidth and delay time can be expressed as a limit to the delay-bandwidth product through the inequality $\Delta\omega\Delta T \leq \kappa$, where $\kappa$ is a dimensionless quantity that generally depends on some general properties of the system, such as its length and maximum permittivity. Various formulations for $\kappa$ have been discussed in the literature for specific types of systems [2], [3], [4]. A short overview of these formulations can be found in Ref. [5].

As with other physical limits, the various delay-bandwidth limits are derived under the assumption of time-invariance. Hence, by incorporating a judiciously designed temporal variation into the system one may be able bypass the delay-bandwidth trade-off. An intuitive method would be to use a time-switched system where the pulse is captured and recirculated within a fiber loop by closing the fiber until the pulse is released [7], [8]. The delay time in such a case can be extended easily beyond conventional limits, as it would depend on the loop length and the number of round trips. However, a large number of round trips would require optical amplifiers to be installed within the loop to compensate for the propagation losses, ultimately increasing the noise level of the system [9]. Moreover, the size of the pulse that can be delayed is restricted by the length of the loop [9]. Another insightful strategy to overcome the delay-bandwidth limit is based on adiabatic tuning of the pulse group velocity $v_\text{g}$ while the pulse is within the delay line (see Figs. 1(a-c)) [2]. An important aspect of this adiabatic modulation is the bandwidth compression of the pulse, which can be easily understood from the corresponding time-dependent dispersion curves of the waveguide mode (Fig. 1(c)). As the waveguide is temporally modulated homogeneously in space while the pulse is within the waveguide, the temporal modulation conserves spatial symmetry; hence, the wavevector components $\Delta k$ of the pulse is conserved. Then, if the temporal modulation is designed such that the dispersion curve of the relevant waveguide mode rotates in the clockwise direction and back to its initial stage during the pulse propagation (as shown in Fig. 1(c)), the bandwidth of the pulse will be compressed and decompressed accordingly to satisfy wavevector conservation. Note that the bandwidth compression enables to bypass the delay-bandwidth limit, since the input pulse with the original bandwidth $\Delta\omega$ can now be slowed down with the delay time $\Delta T'$ which satisfies $\Delta\omega'\Delta T' \leq \kappa$, where $\Delta\omega'$ is the modified bandwidth after the compression. Since $\Delta\omega' < \Delta\omega$, the maximum achievable time delay $\Delta T'$ can be greater than the achievable

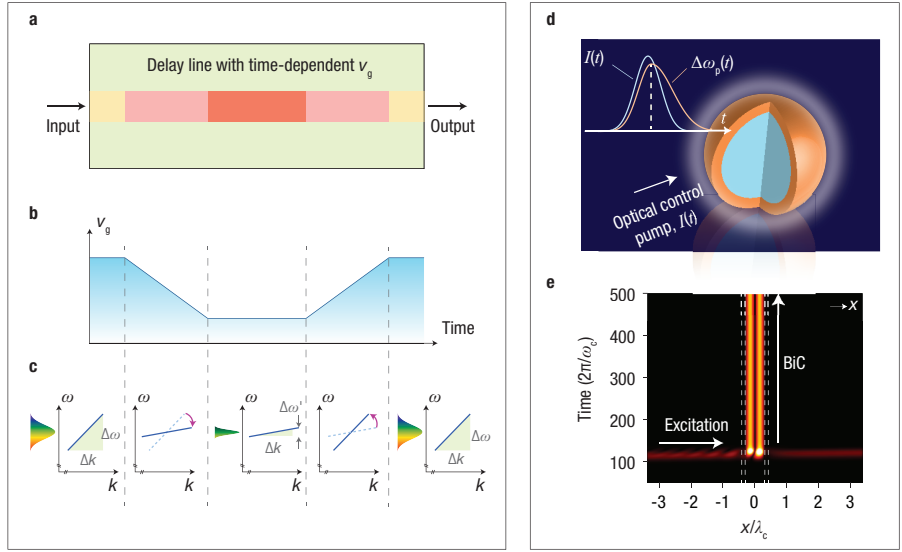

Fig. 1. (a) Schematics of a time-varying delay line. (b) The group velocity $v_g$ of an input pulse is varied in time by dynamically tuning the structural properties of the delay line. (c) The dynamic, spatially homogeneous tuning modifies the dispersion curve of the delay-line waveguide, leading to the adiabatic compression and decompression of the input signal bandwidth $\Delta\omega$, while its wavevector components are conserved. The red arrows show the direction of the temporal adiabatic modulation of the dispersion curves. (d) A compact core-shell scatterer with a time-varying epsilon-near-zero shell can be used to efficiently capture a broadband pulse into a ultra-narrow-bandwidth "bound state in the continuum" (BiC; nonradiating mode of the scatterer) and to bypass the delay-bandwidth limit. The shell permittivity can be varied in time through an optical control pump $I$ that leads to a change $\Delta\omega_p$ in the shell plasma frequency. (e) Electric-field space-time profile showing that a temporally short, broadband, excitation field can be transformed into a long-lived ultra-narrow-bandwidth BiC. $\lambda_c$ is the free space wavelength at the angular frequency $\omega_c$ of the BiC. Parts (d-e) are reproduced with permission from Ref. [6], ACS.

delay if the pulse had experienced the reduced group velocity with no bandwidth compression. Following the pulse delaying, the bandwidth compression can be reversed to reconstruct the original spectrum by time-reversing the initial temporal modulation (Fig. 1(c)). Such adiabatically tuned delaying has been studied, for instance, in coupled-resonator systems [10], [11], [12], [13], [14], [15], [16], photonic crystal waveguides [17], [18], [19], [20], photonic crystal nanocavities [21], ring resonators [22], and recently in spherical core-shell resonant scatterers [6]. For example in Ref. [22] the authors studied a platform composed of a pair of ring resonators coupled to neighboring waveguides. Initially, the resonance frequencies of the two ring resonators are detuned from each other to enable an input pulse to couple from the waveguide into the resonators. During the coupling, the detuning between the resonators is dynamically varied in time using a control pump such that the input pulse experiences a bandwidth compression and the resonator mode gradually becomes decoupled from the waveguides. The lifetime of the resulting supermode that circulates within the two resonators can then be externally controlled through a second control pump and can exceed the delay time dictated by the delay-bandwidth limit of the corresponding time-invariant system. An important limitation of this scheme is that any portion of the pulse that lies outside the two resonators during the temporal modulation gets reflected instead of trapped [23]. To circumvent this issue, in Ref. [6] we have studied a temporally modulated core-shell resonator (see Figs. 1(d,e)), and showed that critical coupling (where the direct reflections destructively interfere with the leakage from the structure) of a free space propagating wave into the resonator mode can be achieved, without any reflections, even for a lossless resonator. Critical coupling in time-modulated platforms was also studied for quasi-monochromatic [24] and chirped [25] waves. While in Ref. [24] this was achieved through temporally modulating the radiative decay rate and in Ref. [25] through modulating the resonance frequency of the resonator, in Ref.

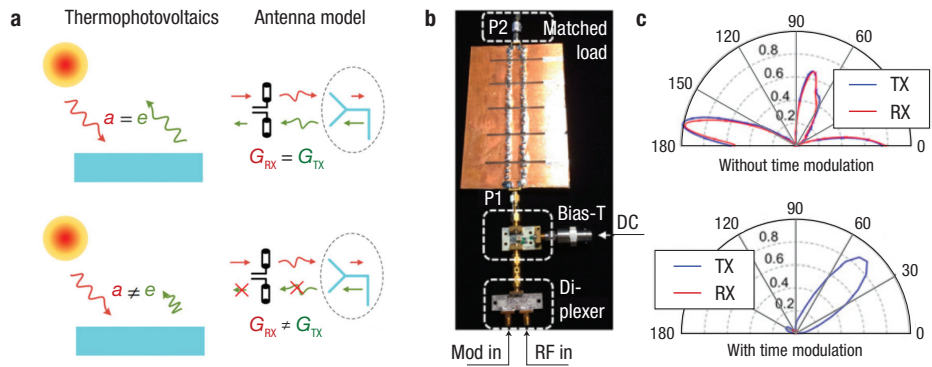

Fig. 2. (a) (upper) A reciprocal system exhibits equal electromagnetic absorption and emission for the same direction and frequency, which limits the ultimate efficiency for various applications ranging from energy harvesting (e.g., photovoltaics) to wireless communication systems. (lower) A nonreciprocal system with unequal absorption and emission for a given direction and frequency, on the other hand, can overcome the typical limitations of such systems, while still satisfying the laws of thermodynamics. (b) Photograph of a fabricated time-modulated nonreciprocal leaky-wave antenna and its feeding network. (c) The antenna radiation patterns for reception (RX) and transmission (TX) are drastically different from each other in the time-modulated case (lower), whereas for the corresponding time-invariant antenna (upper) RX and TX radiation patterns and antenna gain are identical. Reproduced with permission from Ref. [27], National Academy of Sciences.

[6] a Gaussian-like pulse was coupled into an otherwise 'closed' resonator (converting a leaky resonance into a bound state in the continuum), with minimal reflections, by modulating both parameters. Hence, the results in Ref. [6] might provide a link between broadband impedance-matched systems with minimized reflections (Section 2.1 of the main article) and broadband wave trapping/delaying schemes beyond conventional limits. A major practical challenge of the above mentioned works, however, is that the temporal modulation and the incoming pulse need to be synchronized with each other in time. Hence, knowledge about the arrival time of the pulse is required, which might pose challenges in some application scenarios, especially at high frequencies. In this regard, Ref. [26] proposed a scheme based on periodically switched elements that can overcome the delay-bandwidth limit without the need for any pulse-modulation synchronization. These results might likely inspire various other synchronization-free time-varying schemes in different frequency regimes and for different applications, which may pave the way for a new promising direction in this field.

### *1.2. Emission and Absorption; Transmission and Reception*

Due to reciprocity and thermodynamic constraints, a high absorptivity in a certain direction/frequency will also translate into high emissivity in the same direction/frequency, a result known as the Kirchhoff Law of thermal radiation [28]. Hence, as the absorbing material heats up, it will emit some of the absorbed energy via thermal emission, thus reducing the overall amount of harvested energy. Analogous constraints also apply to communication systems, where due to reciprocity an antenna receives and transmits with the same gain in the same direction/frequency (see Fig. 2(a)). Such a constraint might potentially degrade the performance of communication systems as a reciprocal antenna is required to receive its own echoes reflected from the surroundings. In this regard, Ref. [27] investigated how such limitations may be relaxed by breaking reciprocity through time modulation. Specifically, a travelling-wave temporal modulation was applied to a leaky wave antenna to induce asymmetry between the in-coupling and out-coupling process between free space propagating waves and the antenna waveguide mode. This proposal was experimentally demonstrated at microwave frequencies using a leaky transmission line loaded with voltage-dependent varactors that are modulated in time and space to emulate a travelling-wave modulation (see Fig. 2(b)). The resulting antenna radiation measurements (Fig. 2(c)) were shown to confirm that in the absence of any temporal modulation (upper panel of Fig. 2(c)) the radiation patterns

for reception (RX) and transmission (TX) are virtually the same, whereas for the time-modulated case (lower panel of Fig. 2(c)) the RX and TX radiation patterns are drastically different. Hence, it was shown that it is possible to design an antenna that can efficiently radiate in a particular direction, while receiving poorly from all directions. This simple, yet effective, scheme might play an important role in developing compact and efficient communication systems robust against scattering and interference. Similar concepts have also been proposed in other papers to realize efficient energy harvesting systems that can absorb electromagnetic energy incident from a certain direction, without emitting in that direction. Emission will still occur, to satisfy thermodynamics, but in a different direction, where another energy harvesting device can be conveniently placed to absorb this residual radiation. Hence, nonreciprocal absorbers, realized for example through time-modulation, can provide a feasible method to reach the ultimate thermodynamic limit on the efficiency of solar cells – the "Landsberg limit" – which has been theoretically shown to be possible only by breaking reciprocity [28], [29].

### *1.3. Wireless Power Transfer*

In Ref. [30], the authors studied the implications of time-modulated mutual inductances in the context of wireless power transfer, and showed that the trade-off between maximum transferred power and efficiency can be overcome if the modulation pump signal and the source signal are in phase. Specifically, it was noted that, in this case, both the transferred power and efficiency can be increased owing to the parametric gain enabled by modulating the mutual inductance at twice the frequency of the source signal. More broadly, Ref. [30] offers important insights towards understanding the characteristics of time-modulated frequency-dispersive lumped circuit elements, and their usage in various design schemes that may go beyond conventional limits. Possible challenges of the studied schemes could arise due to the internal resistance of the modulation pump and due to the generation of additional harmonics outside the frequency band of interest, both of which would reduce the amount of delivered power to the load.

### *1.4. Camouflaging and Invisibility*

Another interesting application of time-varying electromagnetics is to employ spread-spectrum techniques [31] based on time-switching to reduce the reflection signature of an object for stealth purposes. In this context, for instance, Ref. [32] has demonstrated that by periodically switching between a PEC and a PMC metasurface reflector in a quasi-random fashion, the scattered wave experiences phase discontinuities, resulting in the spreading of the incident spectrum to a broader range. The proposed scheme has also the advantage that a 'friendly' receiver is still able to detect the input signal by demodulating the received scattered signal through a shared temporal modulation key. The proposal was also extended to include space modulations in addition to temporal variations [33]. Such a camouflage technique based on space-time modulated metasurfaces was shown to additionally exhibit spatial spreading of the electromagnetic signal, which could be advantageous for some applications. A similar time-switched metasurface camouflage approach was also studied in Ref. [34]. These time-varying camouflage designs might serve as viable alternatives to conventional stealth technologies and to electromagnetic invisibility cloaks, such as transformation-optics-based or scattering cancellation-based cloaks [35], which suffer from fundamental bounds arising due to their time-invariant and passive nature [36].

### *1.5. Nonlinear Interactions*

Enhanced nonlinear interactions of electromagnetic fields with time-varying systems is another research direction which has been explored, for instance, in Ref. [25]. It was numerically shown that the effective coupling of a spectrally-wide chirped signal to a narrowbanded resonator can enable second harmonic generation beyond the efficiency-bandwidth limit. Another study employed dispersive time-modulated materials to gain direct control over the harmonic generation process to eliminate intermodulation effects [37]. Such harmonic generation schemes may facilitate the development of many practical applications ranging from high-harmonic beam generation [38] to high-resolution imaging [39].

### 1.6. Focusing and Imaging

Increasing the resolution of focusing and imaging systems is highly desirable in many situations. Unfortunately, due to the wave nature of light, the diffraction limit sets a lower bound to the resolvable distance between two objects, at least in far-field schemes using conventional materials. Over the past few decades, various novel artificial materials have been considered to overcome this limitation, such as hyperbolic [40] or negative-index [41] metamaterials, which however suffer from several fundamental issues, such as their sensitivity to losses and imperfections, preventing their practical utilization [42]. Phase-conjugation has emerged as another method to beat the diffraction limit [43], with early proposals for operation at gigahertz frequencies based on nonlinear microwave circuitry controlled through external pumps [44], [45]. Experimental demonstrations of this concept has been reported in the microwave [46] and RF [47] regimes. For optical frequencies, Pendry [48] proposed to use four-wave mixing processes to realize a time-dependent phase-conjugating material, which was later numerically studied using nanoantenna arrays [49] and experimentally studied using an epsilon-near-zero material [50]. Such research efforts may benefit from efficient time-modulation schemes, especially at optical frequencies, and may represent an important step towards future focusing and imaging systems with sub-diffractive resolution.

## Acknowledgements



\end{document}